\documentclass[12pt]{article}
\usepackage{graphicx,epsfig}
\newcommand{\beq}{\begin{equation}}
\newcommand{\eeq}{\end{equation}}
\newcommand{\bqa}{\begin{eqnarray}}
\newcommand{\eqa}{\end{eqnarray}}

\newcommand{\Image}{\mathop{\rm Im}\nolimits}
\newcommand{\sump}{\mathop{{\sum}'}}
\newcommand{\GTSD}{the GT strength distribution\ }
\newcommand{\bld}[1]{\mbox{\boldmath$\mathrm{#1}$}}

\makeatletter
\renewcommand{\@afterindentfalse}{\@afterindenttrue}
\renewcommand{\@seccntformat}[1]{\csname the#1\endcsname.\quad}
\renewcommand{\@cite}[2]{{#1\if@tempswa , #2\fi}}
\makeatother

\begin{document}


\title{On the Fermi and Gamow--Teller strength distributions in medium-heavy mass nuclei}
\author{V.A.~Rodin$^{1),~2)}$, M.H.~Urin$^{2)}$
\thanks{E-mail: vadim.rodin@uni-tuebingen.de, urin@theor.mephi.ru}\\
1.~Institut f\"ur Theoretische Physik der Universit\"at T\"ubingen, Germany\\
2.~Moscow Engineering Physics Institute (State University), Russia}
\date{}

\maketitle

\begin{abstract}
An isospin-selfconsistent approach based on the
Continuum-Random-Phase-Ap\-pro\-xi\-mation (CRPA) is applied to describe
the Fermi and Gamow-Teller strength distributions within a wide
excitation-energy interval. To take into account nucleon pairing in
open-shell nuclei, we formulate an isospin-selfconsistent version of
the proton-neutron-quasiparticle-CRPA (pn-QCRPA) approach by
incorporating the BCS model into the CRPA method. The isospin and
configurational splittings of the Gamow-Teller giant resonance are
analyzed in single-open-shell nuclei. The calculation results
obtained for $^{208}$Bi, $^{90}$Nb, and Sb isotopes are compared with
available experimental data.
\end{abstract}

\section{Introduction}
The Fermi (F) and Gamow--Teller (GT) strength functions in medium-heavy mass
nuclei have been studied for a long time. The subject of studies of the
GT-strength distribution is closely related to the weak probes of nuclei
(single and double $\beta$-decays, neutrino interaction, etc.) as well as to
the direct charge-exchange reactions ((p,n), ($^3$He,t), etc.). The weak
probes deal with the low-energy part of \GTSD (see, e.g.,~[\cite{bm,Suh98}]).
The direct reactions allow to study the distribution in a wide
excitation-energy interval including the region of the GT giant resonance
(GTR) (see, e.g.,~[\cite{Jan93}--\cite{Kras01}] and references therein) and also
the high-energy region (see, e.g.,~[\cite{Wak97,Sak01}]). The discovery of the
isobaric analog resonances (IAR) and the subsequent study of their properties
have allowed one to conclude on the high degree of the isospin conservation in
medium-heavy mass nuclei. It means that the IAR exhausts almost the total
Fermi strength and the rest is mainly exhausted by the isovector monopole
giant resonance (IVMR)~[\cite{bm,Auer83}].

The Random-Phase-Approximation (RPA)-based microscopical studies of \GTSD
started from the consideration of a schematic three-level model taking the
direct, core-polarization, and back-spin-flip transitions into account
~[\cite{ga74}]. One or two weakly-collectivized GT states along with the GTR
were found in the model. Realistic Continuum-Random-Phase-Approximation (CRPA)
calculations for \GTSD were performed in~[\cite{AuerK84}] and~[\cite{Guba89}].
Some of the states predicted in~[\cite{ga74}] were found in~[\cite{Guba89}].
Attempts to describe \GTSD in details within the RPA+Hartree--Fock model have
been undertaken recently in~[\cite{VanG98}]. Unfortunately, the authors of
~[\cite{AuerK84,VanG98}] did not address the following questions: 1) \GTSD in
the high-energy region of the isovector spin-monopole giant resonance (IVSMR);
2) effects of the nucleon pairing; 3) the isospin splitting of the GTR.

Having taken the nucleon pairing into consideration in CRPA calculations, the
authors of~[\cite{Guba89}] predicted the configurational splitting of the GTR in
some nuclei and initiated the respective experimental search for the effect
~[\cite{Jan93}]. In~[\cite{Guba89}], however, the influence of the
particle-particle interaction in the charge-exchange channel was not taken
into account. Along with the pairing interaction in the neutral channel, the
proton-neutron interaction is taken into consideration within the
proton-neutron-quasiparticle-Random-Phase-Approximation (pn-QRPA) to describe
the double $\beta$-decay rates (see, e.g.,~[\cite{Suh98}] and references
therein). Unfortunately, most current versions of the pn-QRPA do not treat the
single-particle continuum that hinders the description of both high-lying IVMR
and IVSMR. In addition, it seems that the question whether the modern versions
of the pn-QRPA comply with the isospin conservation has not been raised yet.

The present paper is stimulated partially by the experimental results of
~[\cite{Jan93}--\cite{Kras01}] and by the intention to overcome
(to some extent) the shortcomings of previous approaches.
As a base we use the isospin-selfconsistent CRPA approach of
~[\cite{Moukh99}--\cite{Rod01}], where direct-decay properties
of giant resonances have been mainly considered.
We pursue the following goals:
\begin{itemize}
\item[1)]
application of the isospin-selfconsistent CRPA approach to describe
the F and GT strength distributions in closed-shell nuclei within a
wide excitation-energy interval including the region of the isovector
monopole and spin-monopole giant resonances;
\item [2)]
formulation of an approximate method to deal with the isospin splitting of
the GTR;
\item[3)]
taking into account the spin-quadrupole part of the particle-hole interaction
for description of GT excitations;
\item[4)]
incorporation of the BCS model into the CRPA method to formulate an
isospin-selfconsistent version of the
proton-neutron-quasiparticle-Continuum-Random-Phase-Approximation (pn-QCRPA)
approach;
\item[5)]
application of the pn-QCRPA approach to describe the F and GT
strength distributions in single-closed-shell nuclei
within a wide excita\-tion-energy interval;
\item[6)]
examination of the GTR configurational splitting effect within the
pn-QCRPA and its impact on the total GTR width.
\end{itemize}
We restrict ourselves to the analysis within the particle-hole subspace, and,
therefore, do not address the question of the influence of
2p--2h configurations (see, e.g.,~[\cite{Bert82}])
as well as the quenching effect (see, e.g.,~[\cite{Arima01}]).
However, we simulate the coupling of the GT states with
many-quasiparticle configurations using an appropriate smearing parameter.

The above-listed goals could be apparently achieved within the pn-QCRPA
approach based on the density-functional method~[\cite{Bor90}]. However, the
authors of~[\cite{Bor90}] focused their efforts mainly on the analysis of
the low-energy part of GT strength distribution relevant for the astrophysical
applications.\footnote{We would emphasise the decisive contribution of
Prof.~S.A.~Fayans in developing such a powerfull approach. We regret much on
his too-early passing away.}

The paper is organized as follows. In Section~2 the basic relationships of the
isospin-selfconsistent CRPA approach are given. They include: description of
the model Hamiltonian, its symmetries and the respective sum rules
(subsections~2.1 and~2.2), the CRPA equations with taking into account the
spin-quadrupole part of the particle-hole interaction for description of the GT
excitations (subsection~2.3); the approximate description of the isospin
splitting of the GTR (subsection~2.4). In Section~3 we extend the CRPA
approach to incorporate the BCS model to describe both the nucleon pairing in
neutral channels and the respective interaction in the charge-exchange
particle-particle channels. The generalized model Hamiltonian, its symmetries
and respective sum rules are given in subsections~3.1 and~3.2. The standard
pn-QCRPA equations are reformulated in terms of radial parts of the transition
density and free two-quasiparticle propagator. That allows us to formulate the
coordinate-space representation for the inhomogeneous system of the pn-CRPA
equations (subsection~3.3). As applied to single-open-shell nuclei, a version
of the pn-QCRPA approach is formulated on the base of the above system
(subsection 3.4). The choice of the model parameters and calculation results
concerned with the F and GT strength distributions in $^{208}$Bi, $^{90}$Nb,
and Sb
 isotopes are presented in Section~4. Summary concerned to the approach
and
 discussion of the calculation results are given in Section~5.

\section{Description of the Fermi and GT strength functions in closed-shell nuclei}
\subsection{Model Hamiltonian}
We use a simple, and at the same time realistic, model Hamiltonian to analyze
the Fermi, ``Coulomb'' (C), and GT strength functions in medium-heavy mass
spherical nuclei. The Hamiltonian consists of the mean field $U(x)$ including
the phenomenological isoscalar part $U_0(x)$ along with the isovector $U_1(x)$
and the Coulomb $U_C(x)$ parts calculated consistently in the Hartree
approximation:
\bqa
&U(x)=U_0(x)+U_1(x)+U_C(x),\label{1.1}\\
&U_0(x)=U_0(r)+U_{so}(x);\ U_1(x)=\displaystyle\frac12 v(r)\tau^{(3)};\ 
U_C(x)=\displaystyle\frac12 U_C(r)(1-\tau^{(3)}).\nonumber
\eqa
Here, $U_0(r)$ and
$U_{so}(x)=U_{so}(r)\bld{\sigma l}$ are the central and spin-orbit parts of the
isoscalar mean field, respectively; $v(r)$ is the symmetry potential. The
potential $U(x)$ determines the single-particle levels with the energies
$\varepsilon_\lambda$ ($\lambda=\pi$ for protons and $\lambda=\nu$ for neutrons,
$\lambda$ is the set of the single-particle quantum numbers,
$(\lambda)=\{lj\}$) and the radial wave functions $r^{-1}\chi_{\lambda}(r)$
along with the radial Green's functions
$(rr')^{-1}g_{(\lambda)}(r,r';\varepsilon)$. On the base of Eq.~(\ref{1.1}),
one can get expressions relating the matrix elements of the single-particle
Fermi ($\tau^{(-)}$) and GT ($\sigma_\mu\tau^{(-)}$) operators ($\sigma_\mu$
are the spherical Pauli matrices):
\bqa
&&t^{(0)}_{(\pi)(\nu)}\{(\varepsilon_\pi-\varepsilon_\nu)(\chi_\pi\chi_\nu)+
v_{\pi\nu}-(U_C)_{\pi\nu}\}=0,\label{1.2}\\
&&t^{(1)}_{(\pi)(\nu)}\{(\varepsilon_\pi-\varepsilon_\nu)(\chi_\pi\chi_\nu)+
v_{\pi\nu}-(U_C)_{\pi\nu}-((\bld{\sigma l})_{(\pi)}-
(\bld{\sigma l})_{(\nu)})(U_{so})_{\pi\nu} \}=0\label{1.2'}
\eqa
with $(\chi_\pi\chi_\nu)=\int \chi_\pi(r)\chi_\nu(r)\,dr$,
$f_{\pi\nu}=\int f(r) \chi_\pi(r) \chi_\nu(r)\,dr$,
$t^{(0)}_{(\pi)(\nu)}=\linebreak\sqrt{2j_\pi+1}\delta_{(\pi)(\nu)}$,
$t^{(1)}_{(\pi)(\nu)}=\frac{1}{\sqrt 3}\langle(\pi)\|\sigma\|(\nu)\rangle$.

We choose the Landau--Migdal forces to describe the particle-hole interaction
~[\cite{Mig83}]. The explicit expression for the forces in the charge-exchange
channel is:
\beq
\hat H_{p\mbox{\it-}h}=2\sum\limits_{a>b}
(F_0+F_1\bld\sigma_a\bld\sigma_b)\tau^{(-)}_a\tau^{(+)}_b
\delta (\bld r_a - \bld r_b) + h.c.\label{1.3}
\eeq
where the intensities of the non-spin-flip and spin-flip parts of this
interaction, $F_0$ and $F_1$ respectively, are the phenomenological parameters.

\subsection{The symmetries of the model Hamiltonian and sum rules.}
The model Hamiltonian $\hat H$ complies with the isospin symmetry provided that
\bqa
&[\hat H, \hat   T^{(-)}]=\hat U^{(-)}_C,\label{1.4} \\
&\hat H=\hat H_0+\hat H_{p\mbox{\it-}h},\ \hat H_0=\sum\limits_a
(T_a+U(x_a)),\ \hat T^{(-)}=\sum\limits_a \tau_a^{(-)},\ \hat U^{(-)}_C=
\sum\limits_a U_C(r_a)\tau_a^{(-)}.\label{1.4'}
\eqa
Using the RPA in the coordinate representation for closed-shell nuclei, one can
get according to Eqs.~(\ref{1.1}), (\ref{1.3})--(\ref{1.4'})
the well-known selfconsistency condition~[\cite{Bir74,Rum98}]:
\bqa
&\displaystyle v(r)=2F_0n^{(-)}(r);\ n^{(-)}(r)=\langle0|\sum_a \tau_a^{(3)}
\delta(\bld r-\bld r_a)|0\rangle=n^{n}(r)-n^{p}(r),\label{1.5} \\
&\displaystyle n^{\beta}(r)=\frac{1}{4\pi r^2}\sum_{\lambda}(2j^\beta_\lambda+1)
n^\beta_\lambda(\chi^\beta_\lambda(r))^2,
\eqa
where $n^{(-)}(r)$ is the neutron excess density, $n^\beta_\lambda$ are the
occupation numbers ($\beta=p,n$). The selfconsistency condition relates
the symmetry potential to the Landau--Migdal parameter $F_0$.

The equation of motion for the GT operator
$\hat Y^{(-)}_\mu=\sum_a (\sigma_\mu)_a \tau_a^{(-)}$ can be derived
within the RPA analogically to Eq.~(\ref{1.4})~[\cite{Rum98}]:
\bqa
&[\hat H, \hat Y^{(-)}_\mu]=\hat U^{(-)}_{\mu},
\label{1.6}\\
&\hat U^{(-)}_{\mu}=\sum_a U_\mu(x_a) \tau_a^{(-)},\ U_\mu(x)=
[U_{so}(x),\sigma_\mu]+(2(F_1-F_0)n^{(-)}(r)+U_C(r))\sigma_\mu.
\label{1.6'}
\eqa

Equations (\ref{1.4}) and (\ref{1.6}) allow one to render
some relationships being useful to
check the calculation results for the strength functions within the RPA.
The strength function corresponding to the single-particle probing operator
$\hat V^{(\mp)}=\sum_a V(x_a) \tau_a^{(\mp)}$ is defined as:
\beq
S^{(\mp)}_V(\omega)=\sum\limits_{s}
\left | \langle s | \hat V^{(\mp)} |0\rangle\right |^2
\delta(\omega-\omega_s)
\label{1.7}
\eeq
with $\omega_s=E_s-E_0$ being the excitation energy of the corresponding
isobaric nucleus measured from the ground state of the parent nucleus.
Using Eqs.~(\ref{1.4}) and (\ref{1.7}) one gets the relationship:
\beq
S^{(\mp)}_F(\omega)=\omega^{-2}S^{(\mp)}_C(\omega).
\label{1.8}
\eeq
Here, the Fermi and ``Coulomb'' strength functions correspond to the probing
operators $\hat T^{(\mp)}$ and $\hat U^{(\mp)}_C$, respectively.

The model-independent non-energy-weighted sum rule ($NEWSR$) for F$(0^+)$ and
GT$(1^+)$ excitations is well-known~[\cite{Ikeda}]:
\beq
(NEWSR)_J=
\langle 0 | \left [ \hat G^{(-)+}_{J\mu},\hat G^{(-)}_{J\mu}\right ] |0\rangle=
(N-Z);\ \hat G^{(-)}_{00}=\hat   T^{(-)};\ \hat G^{(-)}_{1\mu}=\hat Y^{(-)}_\mu.
\label{1.9}
\eeq
The energy-weighted sum rules ($EWSR$)
\beq
(EWSR)_J=
\langle 0 | \left [ \hat G^{(-)+}_{J\mu},
\left [ \hat H, \hat G^{(-)}_{J\mu}\right ] \right ] |0\rangle
\label{1.9'}
\eeq
are rather model-dependent and according to Eqs.~(\ref{1.4}), (\ref{1.6}),
and (\ref{1.6'}) equal to:
\bqa
&&(EWSR)_0= \int U_C(r)n^{(-)}(r)d^3r, \label{1.10}\\
&&(EWSR)_1=(EWSR)_0-\frac43\langle 0 |\hat U_{so} |0\rangle+
2(F_1-F_0)\int (n^{(-)}(r))^2 d^3r
\label{1.11}
\eqa
with $\langle 0 |\hat U_{so} |0\rangle=\sum_\beta\sum_\lambda
n^\beta_\lambda (j^\beta_\lambda-l^\beta_\lambda)l^\beta_\lambda
(l^\beta_\lambda+1)(U_{so})_{\lambda\lambda}$.

According to Eqs.~(\ref{1.4}), (\ref{1.9}), and (\ref{1.10}), the exact
(within the RPA) isospin SU(2) symmetry is realized for the model Hamiltonian
in question in the limit $U_C\to \Delta_C=(N-Z)^{-1}\int U_C(r)n^{(-)}(r)d^3r$.
In this limit the energy $E_A$ and the wave function $|A\rangle$ of the
``ideal'' isobaric analog state (IAS) are $E_A=E_0+\Delta_C$ and
$|A\rangle=(N-Z)^{-1/2}\hat T^{(-)}|0\rangle$, respectively. The ``ideal'' IAS
exhausts 100\% of $(NEWSR)_0$.
In the calculations with the use of a realistic Coulomb mean field, the IAS
exhausts almost 100\% of $(NEWSR)_0$
(with the rest exhausted mainly by the IVMR). Therefore, one can
approximately use the isospin classification of the nuclear states.
Redistribution of the Fermi strength is caused mainly by the Coulomb
mixing of the IAS and the states having the ``normal'' isospin $T_0-1$
($T_0=(N-Z)/2$ is the isospin of both the parent-nucleus ground state and its
analog state). In the present model the mixing is due to the difference
$U_C(r)-\Delta_C$. The Wigner SU(4) symmetry is realized in the limit
$U_{so}\to 0$, $F_1\to F_0$, $U_C\to \Delta_C$. In this limit the energy
$E_{G}$ and the wave function $|G\mu\rangle$ of the ``ideal'' Gamow--Teller
state (GTS) are $E_{G}=E_A$ and
$|G\mu\rangle=(N-Z)^{-1/2}\hat Y^{(-)}_\mu|0\rangle$, respectively.
Redistribution of the GT strength is mainly
due to the spin-orbit part of the mean field.

\subsection{The strength functions within the continuum-RPA}
The distribution of the particle-hole strength can be calculated within the
continuum-RPA making use of the full basis of the single-particle states.
Being based on the above-described model Hamiltonian, the CRPA equations
for calculations of the F\,(C) and GT strength functions
can be derived using the methods of the finite Fermi-system theory
~[\cite{Mig83}]. Let
$\hat V_{JLS\mu}^{(\mp)}=\sum_a V_{JLS\mu}(x_a) \tau_a^{(\mp)}$
be the multipole charge-exchange probing operator leading to excitations with
the angular moment $J$ and parity $\pi=(-1)^L$. Here,
$V_{JLS\mu}(x)=V_{JLS}(r)T_{JLS\mu}(\bld n)$,
$T_{JLS\mu}(\bld n)=\sqrt{4\pi}\sum_{M'M}C^{J\mu}_{LM'SM}Y_{LM'}\sigma_{SM}$
is the irreducible spin-angular tensor operator with
$\sigma_{00}=1,\ \sigma_{1\mu}=\sigma_{\mu}$. In particular, we have
$V_{000}(r)=1\, (U_C(r)),\ {T_{0000}=1}$ and
$V_{101}(r)=1,\ T_{101\mu}=\sigma_{\mu}$ for description of F(C) and GT
excitations, respectively. After separation of the isospin and spin-angular
variables, the strength functions corresponding to the probing operators
$\hat V^{(\mp)}_{JLSM}$ are determined by the following equations:
\bqa
&S_{JLS}^{(\mp)}(\omega)=-\displaystyle\frac1\pi\Image \sum\limits_{L'S'}
\int{V_{JLS}(r)A_{JLS,JL'S'}^{(\mp)}(r,r';\omega)
\tilde V_{JL'S'}^{(\mp)}(r',\omega)\,drdr'},\label{1.12}\\
&\tilde V_{JL'S'}^{(\mp)}(r,\omega)=V_{JLS}(r)\delta_{LL'}\delta_{SS'}+
\displaystyle\frac{2F_{S'}}{4\pi r^2} \sum\limits_{L''S''}
\int{A_{JL'S',JL''S''}^{(\mp)}(r,r';\omega)
\tilde V_{JL''S''}^{(\mp)}(r',\omega)\,dr'}.\label{1.14}
\eqa
Here, $\tilde V^{(\mp)}_{JLS}(r,\omega)$ are the effective radial probing operators,
$F_{S=0,1}$ are the interaction intensities of Eq.~(\ref{1.3}),
$(4\pi rr')^{-2}A_{JLS,JL'S'}^{(\mp)}(r,r';\omega)$
are the radial free particle-hole propagators:
\bqa
&A_{JLS,JL'S'}^{(-)}(r_1,r_2;\omega)=\displaystyle 4\pi\sum\limits_{\nu\pi}
\frac{\langle(\pi)\| T_{JLS} \|(\nu)\rangle \langle(\pi)\| T_{JL'S'} \|(\nu)
\rangle^*}{2J+1}\times\nonumber\\
&\left\{n_{\nu}\chi_{\nu}(r_1)\chi_{\nu}(r_2)
g_{(\pi)}(r_1,r_2;\varepsilon_\nu+\omega)+
n_{\pi}\chi_{\pi}(r_1)\chi_{\pi}(r_2)
g_{(\nu)}(r_1,r_2;\varepsilon_\pi-\omega)\right\}.\label{1.13}
\eqa

The expression for $A^{(+)}$ can be obtained from Eq.~(\ref{1.13}) by the
substitution
$\pi\leftrightarrow\nu$ with $\omega$ being the excitation energy of the
daughter nucleus in the $\beta^+$-channel, measured from the ground state
of the parent nucleus. For $J^\pi=0^+$ excitations ($L=S=0$)
there is only one non-zero propagator $A^{(\mp)}_{000,000}$ and, therefore,
Eqs.~(\ref{1.12}), (\ref{1.14}) have the simplest form. Such a form has been
used explicitly in~[\cite{Gor01a}] for describing IAR and IVMR. For $J^\pi=1^+$
excitations ($S=1$) propagators $A^{(\mp)}_{1L1,1L'1}$ are diagonal
with respect to $S$ and, therefore, Eq.~(\ref{1.14}) is the system of equations
for effective operators $\tilde V_{101}^{(\mp)}$ and $\tilde V_{121}^{(\mp)}$.
As a result, the spin-quadrupole part of the particle-hole interaction
contributes to the formation of the GT strength function, as it follows from
Eqs.~(\ref{1.12}), (\ref{1.14}). The use of only diagonal (with respect to $L$)
propagators $A_{JL1,JL1}^{(\mp)}$ corresponds to the so-called ``symmetric''
approximation. In particular, this approximation was used in
~[\cite{Moukh99,Rod01}] to describe the monopole and dipole spin-flip
charge-exchange excitations. As a rule, the use of the ``symmetric''
approximation leads just to small errors in calculations of the strength
functions $S_{JLS}^{(\mp)}$ in the vicinity of the respective giant resonance
(see, e.g.,~[\cite{Ur92}]). Nevertheless, the detailed description of
the low-energy part of \GTSD appeals to the ``non-symmetric'' approximation.

The F and GT strength functions $S_{J}^{(\mp)}(\omega)$ (hereafter, indexes
$L=0,\linebreak {S=J}$ are omitted) calculated within the CRPA for not-too-high excitation
energies (including the region of IAR and GTR) reveal narrow
resonances corresponding to the particle-hole-type doorway states. Therefore,
the following parametrization holds in the vicinity of each doorway state:
\beq
S_{J}^{(\mp)}(\omega)\simeq  -\frac1\pi\Image
\frac{(r_{J}^{(\mp)})_s}{\omega-\omega_s+\frac i2 \Gamma_s}.\label{1.15}
\eeq
Here, $(r_{J}^{(\mp)})_s$, $\omega_s$, and $\Gamma_s$ are the strength, energy,
and escape width of the doorway state, respectively.
The values of $x_{J}^{tot}=x_{J}^{(-)}-x_{J}^{(+)}=\linebreak
(\int{S_{J}^{(-)}(\omega) d\omega}-
\int{S_{J}^{(+)}(\omega) d\omega})/(NEWSR)_J$,
$y_{J}^{tot}=y_{J}^{(-)}+y_{J}^{(+)}=\linebreak
(\int{\omega S_{J}^{(-)}(\omega) d\omega}+
\int{\omega S_{J}^{(+)}(\omega) d\omega})/(EWSR)_J$
are useful to be compared with unity to check the quality of the calculation
results. Due to the relations given in Eqs.~(\ref{1.7}) and (\ref{1.15})
the ratio of the ``Coulomb'' to the Fermi strengths has to be $\omega_s^2$ for
each doorway state.

\subsection{Isospin splitting of the GT strength distribution.}
Due to the high degree of the isospin conservation in nuclei, the GT states are
classified with the isospin $T_0-1,T_0,T_0+1$. In particular, $T_0$ components
of the GT strength function can be considered as the isobaric analog of the
isovector M1 states in the respective parent nucleus:
\beq
| M, T_0 \rangle=(2T_0)^{-1/2}{\hat T^{(-)}| M1, T_0 \rangle}.
\label{1.16'}
\eeq
According to the definition (\ref{1.7}), the strength function of the $T_0$
components is proportional to the strength function of the isovector M1 giant
resonance in the parent nucleus:
\beq
S_{T_0}^{(-)}(\omega)=(2T_0)^{-1}S_{M1}^{(0)}(\omega'=\omega-\Delta_C).
\label{1.16}
\eeq
Here, $S_{M1}^{(0)}(\omega')$ is the M1 strength function corresponding to the
probing operator $\hat M^{(0)}_\mu=\sum_a (\sigma_\mu)_a \tau_a^{(3)}$ and
depending on the excitation energy in the daughter nucleus. Within the CRPA
the M1 strength function is calculated using the relations, which are
similar to those of Eqs.~(\ref{1.12}), (\ref{1.14}), and can be found, e.g.,
in~[\cite{Rod00}].

According to Eq.~(\ref{1.16}), a noticeable  effect of the isospin
splitting of the GT strength function takes place only for nuclei with a
not-too-large neutron excess. The suppression of the $T_0$ and $T_0+1$
components is the reason why the GT states $|s\rangle$, obtained within the RPA,
are usually assigned the isospin $T_0-1$, although the RPA GT states do not
have a definite isospin. In fact, the states $| s \rangle$ and
$| M, T_0 \rangle$ are non-orthogonal:
\beq
\langle M, T_0| s\rangle=(2T_0)^{-1/2}
\langle M1, T_0|\left [\hat   T^{(+)},\hat Q^{(-)}_s\right] |0\rangle=b_M^s,
\label{1.20}
\eeq
where $\hat Q^{(-)}_s$ is a RPA boson-type operator corresponding to the creation
of a collective GT state $| s\rangle$. Therefore, one has to project
$|s\rangle$ states onto the space of the GT states with the isospin $T_0-1=T_<$
by means of subtraction of the admixtures of $T_0$ states to force the relevant
orthogonality condition:
\beq
| s, T_0-1\rangle=(1-\sum_M (b_M^s)^2)^{-1/2}
(| s\rangle-\sum_M b_M^s|M, T_0\rangle).\label{1.21}
\eeq
This equation is valid under the assumption that the integral relative
strength $x_>=\sum_M x_M$ of the $T_0=T_>$ component is small as compared to
unity. In this case taking into account the $T_0+1$ component is even more
unimportant. We restrict the further analysis by the approximation that there
is the only GT state in the respective RPA calculations exhausting 100\%
of the NEWSR and, therefore, this state can be considered as the ``ideal'' GTS.
Under these assumptions one has:
\beq
b_M^s=(2T_0)^{-1/2}\langle M, T_0|\hat Y^{(-)}|0\rangle=x_M^{1/2};\ x_<=1-x_>.
\label{1.22}
\eeq
Having averaged the exact nuclear Hamiltonian over the state $| s, T_0\rangle$
in the form of Eq.~(\ref{1.21}), one gets:
\beq
\omega_<=x_<^{-1}(x_s\omega_s-\sum_M x_M\omega_M).
\label{1.23}
\eeq
According to the approximate relations (\ref{1.22}) and (\ref{1.23}),
the relative strength and the energy of the $T_0-1$ GTS diminish as
compared with the respective RPA values. The decrease is determined by the
zeroth and first moments of the strength function of the isovector M1 GR in the
parent nucleus:
\beq
x_>=(N-Z)^{-2}\int S^{(0)}_{M1}(\omega')\,d\omega'; \
\sum_M \omega_M x_M=(N-Z)^{-2}\int S^{(0)}_{M1}(\omega')\omega'\,d\omega'.
\label{1.24}
\eeq
These relations again lead to the conclusion that the value\linebreak
$x_>\simeq 2A^{1/3}(N-Z)^{-2}$ is rather small even if the value $(N-Z)$
is not large. Only for nuclei with a minimal neutron excess (a few units) all
three isospin components of the GT strength could have comparable strengths.

\section{The strength functions in open-shell nuclei}
\subsection{The model Hamiltonian}
The interaction $\hat H_{p\mbox{\it-}p}$ in the particle-particle channels has
to be included in the model Hamiltonian for nuclei with open shells, in order
to take the effects of nucleon pairing into consideration. We choose the
interaction in the following separable form (as it is used in the BCS model)
in both the neutral and charge-exchange particle-particle channels with the
total angular momentum and parity of the nucleon pair being
$J^\pi=0^+,1^+\linebreak (L=0, S=J)$:
\beq
\hat H_{p\mbox{\it-}p}=-\frac{1}{2}\sum\limits_{J\mu}G_J\!\!\!\!
\sum\limits_{\scriptstyle\beta=p,n\atop\scriptstyle\beta'=p,n}\!\!
(\hat P^{J\mu}_{\beta\beta'})^+ \hat P^{J\mu}_{\beta\beta'}\ .\label{2.1}
\eeq
Here, $G_{J=0,1}$ are the intensities of the particle-particle interaction,
$P^{J\mu}$ is the annihilation operator for the nucleon pair:
\bqa
&&\hat P^{J\mu}_{\beta\beta'}=\sum\limits_{\lambda\lambda'}
(\chi_\lambda^\beta \chi_{\lambda'}^{\beta'})
P^{J\mu}_{\beta\beta',\lambda\lambda'}\ ;\
P^{J\mu}_{\beta\beta',\lambda\lambda'}=
t^{(J)}_{(\lambda)(\lambda')}\sum\limits_{mm'}
(J\mu|j_\lambda^\beta m j_{\lambda'}^{\beta'} m') a^{\beta'}_{\lambda' m'}
a^{\beta}_{\lambda m}\ ,\label{2.2}
\eqa
where $a_{\lambda m}(a_{\lambda m}^+)$ is the annihilation (creation)
operator of the nucleon in the state with the quantum numbers $\lambda m$
($m$ is the projection of the particle angular momentum).
The interaction (\ref{2.1}) preserves
the isospin symmetry of the model Hamiltonian.

We use the Bogolyubov transformation to describe the nucleon pairing in the
neutral channels in terms of quasiparticle creation (annihilation) operators
$\alpha_{\lambda m}^+(\alpha_{\lambda m})$ (see, e.g.,~[\cite{Sol}]). As a
result, we get the following model Hamiltonian to describe F$\,$(C) and GT
excitations in the $\beta^-$-channel within the quasiboson version of the
pn-QRPA:
\beq
\hat H=\hat H_{0}+\hat H_{p\mbox{\it-}h}+\hat H_{p\mbox{\it-}p};\
\hat H_{0}=\sum\limits_{\beta=p,n}\sum\limits_{\lambda m}
E^\beta_\lambda (\alpha^\beta_{\lambda m})^+\alpha^\beta_{\lambda m}-
\frac12 (\mu_p-\mu_n)\hat T^{(3)}.\label{2.4}
\eeq
Here, $E^\beta_\lambda=\sqrt{(\xi^\beta_\lambda)^2+\Delta_\beta^2}$
is the quasiparticle energy, $\xi^\beta_\lambda=
\varepsilon^\beta_\lambda-\mu_\beta$, $\mu_\beta$ and $\Delta_\beta$
are the chemical potential and the energy gap, respectively,
which are determined from the BCS-model equations:
\beq
N_\beta=\sum_\lambda
(2j^\beta_\lambda+1) (v^\beta_\lambda)^2;\ \ \
\Delta_\beta=G_0 \sum_\lambda
(2j^\beta_\lambda+1) u^\beta_\lambda v^\beta_\lambda
\label{2.5}
\eeq
with $v^2_\lambda=\frac{1}{2}(1-\frac{\xi_\lambda}{E_\lambda})$
and $u^2_\lambda=1-v^2_\lambda$.

The total interaction Hamiltonian in both particle-hole (\ref{1.3}) and
particle-particle (\ref{2.1}) channels can be expressed in terms of
the quasiparticle (pn)-pair creation and annihilation operators which obey
approximately the bosonic commutation rules:
\beq
A^{J\mu}_{\pi\nu}=\sum\limits_{mm'}
(J\mu|j_\pi mj_{\nu} m') \alpha_{\nu m'} \alpha_{\pi m};\ \ \
[A^{J'\mu'}_{\pi'\nu'},(A^{J\mu}_{\pi\nu})^+]=\delta_{\pi\pi'}
\delta_{\nu\nu'}\delta_{JJ'}\delta_{\mu\mu'}.
\label{2.6}
\eeq
The explicit expressions for the interactions are:
\bqa
&&\hat H_{p\mbox{\it-}h}=\sum\limits_{J\mu}\frac{2F_J}{4\pi}
\sum_{\pi\nu\pi'\nu'}(\chi_\pi \chi_\nu \chi_{\pi'} \chi_{\nu'})
(Q^{J\mu}_{\pi\nu})^+Q^{J\mu}_{\pi'\nu'},\label{2.7}\\
&&\hat H_{p\mbox{\it-}p}=-\sum\limits_{J\mu}G_J\sum_{\pi\nu\pi'\nu'}
(\chi_\pi \chi_\nu)(\chi_{\pi'} \chi_{\nu'})
(P^{J\mu}_{\pi\nu})^+P^{J\mu}_{\pi'\nu'},\label{2.8}
\eqa
where $Q^{J\mu}_{\pi\nu}=t^{(J)}_{(\pi)(\nu)}(u_\pi v_\nu A^{J\mu}_{\pi\nu}+
v_\pi u_\nu (\tilde A^{J\mu}_{\pi\nu})^+)$,
$P^{J\mu}_{\pi\nu}=t^{(J)}_{(\pi)(\nu)}(u_\pi u_\nu A^{J\mu}_{\pi\nu}
-v_\pi v_\nu (\tilde A^{J\mu}_{\pi\nu})^+)$,
$\tilde A^{J\mu}_{\pi\nu}=(-1)^{J+\mu} A^{J\;-\mu}_{\pi\nu}$,
$(\chi_\pi \chi_\nu \chi_{\pi'} \chi_{\nu'})=
\int \chi_\pi \chi_\nu \chi_{\pi'} \chi_{\nu'} r^{-2}dr$.

\subsection{The symmetries of the model Hamiltonian and sum rules.}
Due to the isobaric invariance of the particle-particle interaction (\ref{2.1}),
the Eq.~(\ref{1.4}) still holds and leads exactly to the same selfconsistency
condition of Eq.~(\ref{1.5}) with the only difference, that the proton and
neutron densities are determined with account for the particle redistribution
caused by the nucleon pairing:
\beq
n^{(\beta)}(r)=\frac{1}{4\pi r^2}\sum_{\lambda}(2j^\beta_\lambda+1)
(v^\beta_\lambda)^2(\chi^\beta_\lambda(r))^2. \label{2.9}
\eeq
The direct realization of the Eq.~(\ref{1.4}) within the pn-QRPA with making
use of Eqs.~(\ref{2.4})--(\ref{2.8}) and
$\hat  T^{(-)}=\sum_{\pi\nu} (\chi_\pi \chi_\nu) (Q^{00}_{\pi\nu})^+$
leads also to the selfconsistensy condition of Eq.~(\ref{1.5}) provided that
the full basis of the single-particle states for neutron and proton subsystems
is used along with Eq.~(\ref{1.2}) for the radial overlap integrals for the
proton and neutron wave functions. The consistent mean Coulomb field $U_C(r)$
and $(EWSR)_0$ of Eq.~(\ref{1.10}) are determined by the proton and neutron
excess densities of Eq.~(\ref{2.9}), respectively.

The Eq.~(\ref{1.8}) relating the Fermi and ``Coulomb'' strength functions holds also within
the pn-QRPA if the all above-mentioned conditions are fulfilled. In particular,
the use of a truncated basis of the single-particle states within the BCS model leads
to a unphysical violation of the isospin symmetry. In such a case the degree of
violating Eq.~(\ref{1.8}) can be considered as a measure of the violation.

The equation of motion for the GT operator $\hat Y^{(-)}_{\mu}$ with taking the
nucleon pairing into consideration is somewhat modified as compared to
Eqs.~(\ref{1.6}), (\ref{1.6'}). According to Eqs.~(\ref{2.4})--(\ref{2.8}) and
making use of the full basis of the single-particle states along with
Eq.~(\ref{1.2}) for the radial overlap integrals we get within the pn-QRPA
($\hat  Y^{(-)}_\mu=\sum_{\pi\nu}(\chi_\pi \chi_\nu)(Q^{1\mu}_{\pi\nu})^+$):
\beq
[\hat H, \hat Y^{(-)}_{\mu}]=(\hat U^{(-)}_{\mu})_{p\mbox{\it-}h}+
(\hat U^{(-)}_{\mu})_{p\mbox{\it-}p}\ .\label{2.10}
\eeq
The operator $(\hat U^{(-)}_{\mu})_{p\mbox{\it-}h}$ is defined by the
expression~(\ref{1.6'}) in which both the neutron excess and proton densities
are used according to Eq.~(\ref{2.9}). The expression for
$(\hat U^{(-)}_{\mu})_{p\mbox{\it-}p}$
has the form:
\beq
(\hat U^{(-)}_{\mu})_{p\mbox{\it-}p}=
\frac{G_0-G_1}{G_0}(\Delta_n (\hat P^{1\mu}_{pn})^+ -
\Delta_p \tilde{\hat P}^{1\mu}_{pn}).\label{2.11}
\eeq
According to Eqs.~(\ref{1.6'}), (\ref{2.10}), and (\ref{2.11}) the expression
for $(EWSR)_1$ consists of two terms, the former, $(EWSR)_1^{p\mbox{\it-}h}$,
coinciding with Eq.~(\ref{1.11}) with the nucleon densities and the
occupation numbers appropriately modified by the nucleon pairing, and
the latter being due to the nucleon pairing only:
\beq
(EWSR)_1^{p\mbox{\it-}p}=\frac{G_0-G_1}{G_0}\frac{\Delta_n^2 + \Delta_p^2}{G_0}.
\label{2.12}
\eeq
In the SU(4)-symmetry limit one has the equality $(EWSR)_1=(EWSR)_0$, which
holds also for double-closed-shell nuclei.

\subsection{The pn-QRPA equations.}
The system of the homogeneous equations for the forward and backward amplitudes
{$X^{J}_{\pi\nu}(s)=\langle s,J\mu | \left ( A^{J\mu}_{\pi\nu} \right )^+ |0\rangle$}
and $Y^{J}_{\pi\nu}(s)=\langle s,J\mu |\tilde A^{J\mu}_{\pi\nu} |0\rangle$,
respectively, is usually solved to calculate the energies $\omega_s$ and the
wave functions $|s,J\mu\rangle$ of the isobaric nucleus within the quasiboson
version of the pn-QRPA (see, e.g.,~[\cite{Suh98}]). In particular, the system
of the equations for the amplitudes follows from the equations of motion for
the operators $A^+$ and $\tilde A$ making use of the Hamiltonian
(\ref{2.4}),(\ref{2.7}),(\ref{2.8}). Instead, we rewrite the system in
equivalent terms of the elements $r^{-2}\varrho^{J}_i(s,r)$ of the radial
transition density. The elements are determined by the amplitudes
$X(s)$ and $Y(s)$ as follows:
\bqa
&&
\varrho^{J}_i(s,r)=\sum\limits_{\pi\nu}t^{(J)}_{(\pi)(\nu)}\chi_\pi(r)\chi_\nu(r)
\langle s,J\mu | \left ( R^{J\mu}_{\pi\nu} \right )_i |0\rangle, \ \ i=1,2,3,4;
\label{2.13}\\
&&
\left ( R^{J\mu}_{\pi\nu} \right )_1=\left ( Q^{J\mu}_{\pi\nu} \right )^+,\ \
\left ( R^{J\mu}_{\pi\nu} \right )_2=\tilde Q^{J\mu}_{\pi\nu},\ \
\left ( R^{J\mu}_{\pi\nu} \right )_3=\tilde P^{J\mu}_{\pi\nu},\ \
\left ( R^{J\mu}_{\pi\nu} \right )_4=\left ( P^{J\mu}_{\pi\nu} \right )^+,
\label{2.13'}
\eqa
where the operators $P$ and $Q$ are defined after Eqs.~(\ref{2.7}), (\ref{2.8}).
According to the definition (\ref{2.13}), the elements
$\varrho_1,\varrho_2,\varrho_3,\varrho_4$ can be called, respectively,
the particle-hole, hole-particle, hole-hole and particle-particle components of
the transition density, which can be generally considered as a 4-dimensional
vector: $\{ \varrho^{J}_i\}$. In particular, the particle-hole strength of the
state $|s,J\mu\rangle$ corresponding to a probing operator
$\hat V^{(-)}_{J\mu}$ is determined by the element $\varrho^{J}_1$:
$(r^{(-)}_{J})_s=\left (4\pi\int \varrho^J_1(s,r) V_{J}(r)\, dr\right )^2$
(compare to Eq.(\ref{1.15})). The pn-QRPA system of equations for the elements
$\varrho_i^{J}$ is the following (hereafter the ``symmetric'' approximation is
only considered):
\bqa
&&\varrho^{J}_i(s,r)=\frac{1}{4\pi}\sum\limits_{k} \int
A^{(-)}_{J,ik}(r,r_1;\omega=\omega_s)F^{J}_k(r_1,r_2)\varrho^{J}_k(s,r_2)
\, dr_1dr_2, \label{2.14}\\
&&F^{J}_1(r_1,r_2)=F^{J}_2(r_1,r_2)=2F_J\frac{\delta(r_1-r_2)}{r_1r_2},\
F^{J}_3(r_1,r_2)=F^{J}_4(r_1,r_2)= - 4\pi G_J, \label{2.15}\\
&&A^{(-)}_{J,ik}(r_1,r_2;\omega)=\sum\limits_{\pi\nu}
\left (t^{(J)}_{(\pi)(\nu)}\right )^2
\chi_\pi(r_1)\chi_\nu(r_1)\chi_\pi(r_2)\chi_\nu(r_2)
A^{(-)}_{\pi\nu,ik}(\omega), \label{2.16}\\
&& A^{(-)}_{\pi\nu,11}=\frac{u^2_\pi v^2_\nu}{\{ - \}} - \frac{u^2_\nu v^2_\pi}{\{ + \}}\ ,\ \
A^{(-)}_{\pi\nu,12}=u_\pi v_\pi v_\nu u_\nu(\frac{1}{\{ - \}} - \frac{1}{\{ + \}}),
\nonumber\\
&&A^{(-)}_{\pi\nu,14}=u_\pi v_\pi (\frac{v^2_\nu}{\{ - \}} + \frac{u^2_\nu}{\{ + \}})\ ,\ \
A^{(-)}_{\pi\nu,13}=u_\nu v_\nu (\frac{u^2_\pi}{\{ - \}} + \frac{v^2_\pi}{\{ + \}}),
\nonumber\\
&&A^{(-)}_{\pi\nu,22}=\frac{v^2_\pi u^2_\nu}{\{ - \}} - \frac{v^2_\nu u^2_\pi}{\{ + \}}\ ,\ \
A^{(-)}_{\pi\nu,23}=u_\pi v_\pi (\frac{u^2_\nu}{\{ - \}} + \frac{v^2_\nu}{\{ + \}}),
\nonumber\\
&&A^{(-)}_{\pi\nu,24}=u_\nu v_\nu (\frac{v^2_\pi}{\{ - \}} + \frac{u^2_\pi}{\{ + \}})\ ,\ \
A^{(-)}_{\pi\nu,33}=\frac{u^2_\pi u^2_\nu}{\{ - \}} - \frac{v^2_\nu v^2_\pi}{\{ + \}},
\nonumber\\
&& A^{(-)}_{\pi\nu,44}=\frac{v^2_\pi v^2_\nu}{\{ - \}} - \frac{u^2_\nu u^2_\pi}{\{ + \}}\ ,\ \
A^{(-)}_{\pi\nu,ik}=A^{(-)}_{\pi\nu,ki} ;\ \ A^{(-)}_{\pi\nu,34}=A^{(-)}_{\pi\nu,12},
\nonumber
\eqa
where $\{ - \}=\omega^{(-)}-E_\pi-E_\nu$, $\{ + \}=\omega^{(-)}+E_\pi+E_\nu$,
$\omega=\omega^{(-)}+\mu_p-\mu_n$ gives the excitation energy with respect to
the mother-nucleus ground-state energy. All actual calculations in present work
are performed in terms of $\omega$, whereas the experimental energy difference
between the ground states of daughter and mother nuclei is used to represent
the results in terms of daughter-nucleus excitation energy $E_x$.

According to Eq.~(\ref{2.14}), schematically represented as
$\varrho^{J}_i=\sum_k A^{(-)}_{J,ik}(\omega=\omega_s) F^J_k \varrho^J_k$,
the $4\times 4$ matrix
$(4\pi r_1^2r_2^2)^{-1}A^{(-)}_{J,ik}(r_1,r_2;\omega)$ is the radial part
of the free two-quasiparticle propagator, whereas the quantities
$v^{J}_k(s,r_1)=\int F^{J}_k(r_1,r_2)\varrho^{J}_k(s,r_2)\,dr_2$ are the
elements of the radial transition potential. To describe the excitations in
$\beta^+$-channel one should use the same system (\ref{2.14}) with $A^{(-)}$
superseded by $A^{(+)}$. The expression for the particle-hole propagator
$A^{(+)}$ can be obtained from Eqs.~(\ref{2.16}) by the substitutions
$\nu \leftrightarrow\pi,\,n \leftrightarrow p$.
In the case of neglecting nucleon pairing ($G_J=0$), the system of equations
(\ref{2.14}) decouples and $A^{(-)}_{J,11}$ becomes determined by
Eq.~(\ref{1.13}) taken in the ``symmetric'' approximation (it can be shown
easily using the spectral expansion for the radial Green's functions
$g_{(\lambda)}(r,r';\varepsilon)$).

\begin{figure}[ht]
\begin{center}
\epsfig{file=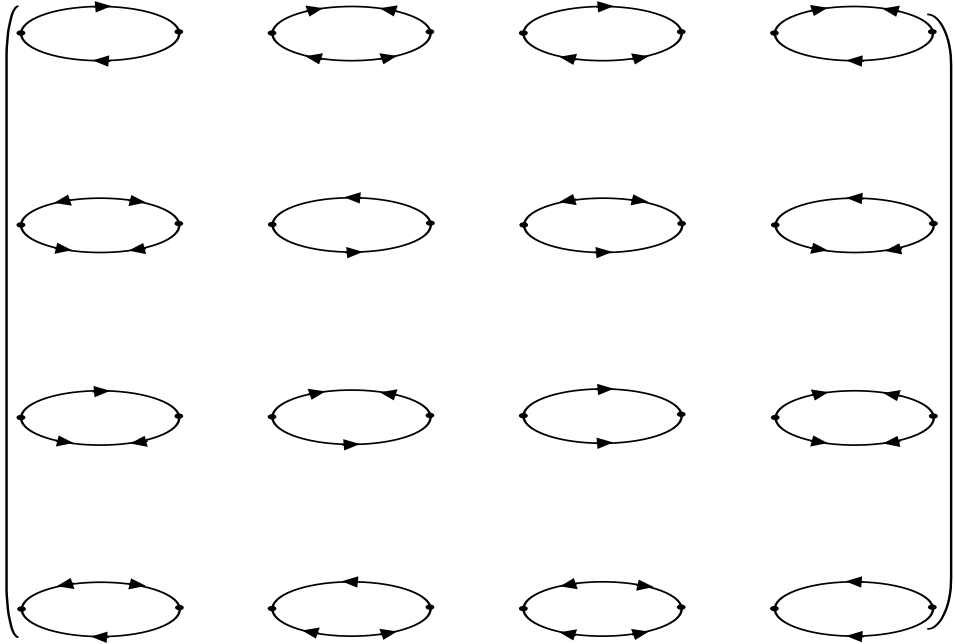,height=5in,width=5in}
\caption{Diagrams corresponding to elements $A^{(-)}_{\pi\nu,ik}$ of the free
two-quasiparticle propogator of Eq.~(\ref{2.16}). Upper (lower) lines
correspond to propagation of proton (neutron) quasiparticles.}
\label{fig1}
\end{center}
\end{figure}

The expression for the elements of the free two-quasiparticle propagator
(\ref{2.16}) can be obtained making use of the regular and anomalous
single-particle Green's functions for Fermi-systems with nucleon pairing in
analogue way as it was done in monograph~[\cite{Mig83}] to describe the
Fermi-system response to a single-particle probing operator acting in the
neutral channel. Namely, the matrix $A$ can be depicted as a set of the
diagrams shown in Fig.~\ref{fig1}. To illustrate this statement, let us
consider the system of equations for the elements of the transition potential
$v_i$. The system follows from Eq.~(\ref{2.14}) and can be schematically
represented as $v_i=\sum_k F_i A^{(-)}_{ik}(\omega=\omega_s) v_k\ .$

Using expression (\ref{2.16}) for the free two-quasiparticle propagator, one
can obtain the system of the inhomogeneous equations of the pn-QRPA to
calculate the strength functions (\ref{1.7}) with taking the full basis of the
single-particle states into consideration for the particle-hole channel.
The system has the form similar to that of Eqs.~(\ref{1.12}), (\ref{1.14})
taken in the ``symmetric'' approximation:
\bqa
&\displaystyle S_{J}^{(\mp)}(\omega)=-\frac1\pi\Image \sum\limits_{i}\int
V_J(r) A_{J,1i}^{(\mp)}(r,r';\omega)\tilde V_{J,i}^{(\mp)}(r',\omega)\, drdr',
\label{2.18}\\
&\displaystyle \tilde V_{J,i}^{(\mp)}(r,\omega)=V_{J}(r)\delta_{i1}+
\frac{1}{4\pi r^2}\sum\limits_{k} \int F^J_i(rr_1)
A_{J,ik}^{(\mp)}(r_1,r_2;\omega)\tilde V_{J,k}^{(\mp)}(r_2,\omega)\, dr_1dr_2.
\label{2.19}
\eqa
The same as in case without pairing, the residues $(r^{(\mp)}_J)_s$ in poles
of the strength functions (see Eq.~(\ref{1.15})) coincide with the
particle-hole strengths of the pn-QRPA excitations corresponding to the probing
operators $\hat V^{(\mp)}_{J\mu}$.

\subsection{Strength functions in single-open-shell nuclei.}
We restrict the further consideration to the case of the single-open-shell
nuclei. For this case the system of equations (\ref{2.19}) for the radial
effective operators $\tilde V^{(\mp)}_{J,i}(r,\omega)$ is noticeably simplified.
Let us consider for definiteness the excitations in the $\beta^-$-channel for a
nucleus in which only the proton pairing is realized ($\Delta_n=0$).
Then, according to Eq.~(\ref{2.16}), the elements
$A^{(-)}_{J,12},A^{(-)}_{J,13},A^{(-)}_{J,24},A^{(-)}_{J,34}$ of the free
two-quasiparticle propagator equal to zero and the system (\ref{2.19})
is reduced to the system of equations only for 2 radial effective operators
$\tilde V^{(-)}_{J,1}$ and $\tilde V^{(-)}_{J,4}$. The elements
$A^{(-)}_{J,11},A^{(-)}_{J,14},A^{(-)}_{J,44}$ entering this system can be
rewritten in the form which allows one to take the full basis of the
single-particle states in the particle-hole channel for both subsystems.
To get this form one should put $E_\pi=|\varepsilon_\pi-\mu_\pi|$ for
$E_\pi>k\Delta_p$ ($k\gg 1$) and use the spectral expansion for the radial
single-particle Green's functions in Eq.~(\ref{2.16}). As a result, we come to
the following representation for $A^{(-)}_{J,11}$:
\bqa
&\displaystyle A^{(-)}_{J,11}(r_1,r_2;\omega)=\sum\limits_{\nu(\pi)}\left
(t^{(J)}_{(\pi)(\nu)}\right )^2
n_\nu \chi_\nu(r_1)\chi_\nu(r_2) g_{(\pi)}(r_1,r_2;\varepsilon_\nu+\omega) +
\label{2.20}\\
&\displaystyle \sum\limits_{(\nu)\pi}\left (t^{(J)}_{(\pi)(\nu)}\right )^2
v^2_\pi \chi_\pi(r_1)\chi_\pi(r_2) g_{(\nu)}(r_1,r_2;\mu_\pi-E_\pi-\omega) +
\nonumber\\
&\displaystyle \sump\limits_{\nu\pi}\left (t^{(J)}_{(\pi)(\nu)}\right )^2
n_\nu \chi_\nu(r_1)\chi_\nu(r_2)\chi_\pi(r_1)\chi_\pi(r_2)\times\nonumber\\
&\displaystyle\left \{
\frac{u^2_\pi}{\omega-\mu_p-E_\pi+\varepsilon_\nu}+
\frac{v^2_\pi}{\omega-\mu_p+E_\pi+\varepsilon_\nu}-
\frac{1}{\omega-\varepsilon_\pi+\varepsilon_\nu}
\right \},
\nonumber
\eqa
where $\sump_{\nu\pi}$ denotes the sum, which is taken over the proton states
with $E_\pi < k\Delta_p$ only. For the consistency of the consideration,
the same truncation should be used in realization of the BCS model according
to the second of Eqs.~(\ref{2.5}). The corresponding expressions for the
elements $A^{(-)}_{J,14}$ and $A^{(-)}_{J,44}$ are:
\bqa
&\displaystyle A^{(-)}_{J,14}(r_1,r_2;\omega)=-\sump\limits_{(\nu)\pi}
\left (t^{(J)}_{(\pi)(\nu)}\right )^2
u_\pi v_\pi \chi_\pi(r_1)\chi_\pi(r_2) g_{(\nu)}(r_1,r_2;\mu_\pi-E_\pi-\omega)+
\label{2.21}\\
&\displaystyle \sump\limits_{\nu\pi} \left (t^{(J)}_{(\pi)(\nu)}\right )^2
n_\nu \chi_\nu(r_1)\chi_\nu(r_2)\chi_\pi(r_1)\chi_\pi(r_2)
\left \{
\frac{1}{\omega-\mu_p-E_\pi+\varepsilon_\nu}-
\frac{1}{\omega-\mu_p+E_\pi+\varepsilon_\nu}
\right \}, \nonumber\\
&\displaystyle A^{(-)}_{J,44}(r_1,r_2;\omega)=\sump\limits_{(\nu)\pi}
\left (t^{(J)}_{(\pi)(\nu)}\right )^2
u_\pi^2 \chi_\pi(r_1)\chi_\pi(r_2) g_{(\nu)}(r_1,r_2;\mu_\pi-E_\pi-\omega)+
\label{2.22}\\
&\displaystyle \sump\limits_{\nu\pi} \left (t^{(J)}_{(\pi)(\nu)}\right )^2
n_\nu \chi_\nu(r_1)\chi_\nu(r_2)\chi_\pi(r_1)\chi_\pi(r_2) \left \{
\frac{v^2_\pi}{\omega-\mu_p-E_\pi+\varepsilon_\nu}+
\frac{u^2_\pi}{\omega-\mu_p+E_\pi+\varepsilon_\nu}
\right \} .
\nonumber
\eqa
The equations (\ref{2.18})--(\ref{2.22}) realize a continuum version of the
pn-QRPA (pn-QCRPA) on the truncated basis of the BCS problem in
proton-open-shell nuclei.

To go to the case when only the neutron pairing is realized, the equations
(\ref{2.20})--(\ref{2.22}) should be changed by the substitution
$\pi\leftrightarrow\nu,\ p\leftrightarrow n,\ \omega\to -\omega$.
As before, the expression for the elements of the free two-quasiparticle
propagator $A^{(+)}$ can be obtained from the corresponding elements
$A^{(-)}$ by the substitution $\pi\leftrightarrow\nu,\ p\leftrightarrow n$.

\section{Calculation results}
\subsection{Choice of the model parameters}
Parametrization of the isoscalar part of the mean field $U_0(x)$ along with
the values of the parameters used in the calculations have been described
in details in~[\cite{Gor01b}]. The dimensionless intensities $f_J$ of
the Landau-Migdal forces of Eq.~(\ref{1.3}) are chosen as usual:
$F_J=f_J\cdot300$ MeV\,fm$^3$.
The value $f_0=f'=1.0$ of the parameter $f'$ determining the symmetry potential
according to Eq.~(\ref{1.5}) is also taken from the~[\cite{Gor01b}], where the
experimental nucleon separation energies have been satisfactorily described
for closed-shell subsystems in a number of nuclei. The value of
the dimensionless intensity $f_1=g'$ of the spin-isovector part of
the Landau-Migdal forces $g'=0.8$ is chosen to reproduce within the CRPA
the experimental energy of the GTR in $^{208}$Pb parent nucleus~[\cite{Moukh99}].

The strength of the monopole particle-particle interaction $G_0$ is chosen to
reproduce the experimental pairing energies (actually, we identify the pairing
energy with $\Delta$, obtained by solving the BCS-model equations (\ref{2.5})).
The summation in the second of the equations is limited by the interval of
7 MeV above and below the Fermi level. The same truncation is used in the
expressions like (\ref{2.20})--(\ref{2.22}) for the elements of the free
two-quasiparticle propagator. Comparing the calculated value of the nucleon
separation energy $B_\beta^{calc}\simeq-\mu_\beta+\Delta_\beta$ with the
corresponding experimental one can be considered as a test of the used version
of the BCS model. The parameters of the BCS model along with $B_\beta^{calc}$
and $B_\beta^{exp}$ are listed in the Table~\ref{tab1}. Note also, that the
strength of the spin-spin particle-particle interaction $G_1$ is chosen equal
to $G_0$, which seems close to the realistic ones used in literature. Such a
choice allows us to simplify the calculation control using the $(EWSR)_1$,
because the ``particle-particle'' part of this sum rule goes to zero in
accordance with Eq.~(\ref{2.12}). Another reason for the choice is the ``soft''
spin-isospin SU(4)-symmetry, which is roughly realized in nuclei
(see, e.g.,~[\cite{Gap92}]).

\subsection{F\,(C) strength functions}
The F(C) strength functions have been calculated within the CRPA for the
$^{208}$Pb parent nucleus according to Eqs.~(\ref{1.12}), (\ref{1.14}) and
within the pn-QCRPA for $^{90}Zr$ and tin isotopes according to
Eqs.~(\ref{2.18}), (\ref{2.19}) or their modifications. The ``cut-off''
parameter $k=3$ was used in calculations of the radial propagators of
Eqs.~(\ref{2.20})--(\ref{2.22}). All above-mentioned equations have been taken
for $J=0$. The following characteristics of the F strength distribution have
been deduced from the calculated F strength functions: the IAR energy $E_{IAR}$;
the mean energy $\bar E_{IVMR}^{(-)}$ of the IVMR$^{(-)}$; relative Fermi
strength $x$ for 3 energy intervals: the vicinity of the IAR, IVMR$^{(-)}$, and
IVMR$^{(+)}$; relative energy-weighted Fermi strength $y$ for these 3 energy
regions. As for the low-energy interval, for all nuclei in question the values
of $x$ are below 0.1\%
(except for $^{90}$Zr, where $x\simeq$0.2\%).
The value of $(EWSR)_0$ entering the definition of $y$ has been calculated
according to Eq.~(\ref{1.10}) with the nucleon densities modified by the
nucleon pairing. All the characteristics of the F strength distribution are
listed in the Table~\ref{tab2} along with the experimental IAR energies.
To check the selfconsistency of the model, the parameter
$z=\left|\frac{\int S^{(-)}_C(\omega)\, d\omega}
{\int S^{(-)}_F(\omega)\omega^2\, d\omega}-1 \right|$
has also been calculated.

\subsection{GT strength functions}
The GT strength functions have been calculated within the CRPA for the
$^{208}$Pb parent nucleus according to Eqs.~(\ref{1.12}), (\ref{1.14}) and
within the pn-QCRPA for $^{90}$Zr and tin isotopes according to
Eqs.~(\ref{2.18}), (\ref{2.19}) or their modifications taken for $J=1$.
>From the calculated GT strength functions the relative GT strength $x$ and
relative energy-weighted GT strength $y$ for 4 energy intervals (low-energy,
the vicinity of the GTR, high-energy, and the vicinity of the IVSMR$^{(+)}$)
have been deduced. The value of $(EWSR)_1^{p\mbox{\it-}h}$ entering the
definition of $y$ has been calculated according to Eq.~(\ref{1.11}) with the
nucleon densities and the occupation numbers modified by the nucleon pairing.
The value of $(EWSR)_1^{p\mbox{\it-}p}$ defined according to Eq.~(\ref{2.12})
equals zero because of putting $G_1=G_0$ in our calculations. The values of the
parameters $x$ and $y$ are listed in the Table~\ref{tab3} along with the
calculated energies of the GTR. The values of the relative strengths
$x_s=r_s/(N-Z)$ of the GT doorway states calculated according to
Eq.~(\ref{1.15}) with and without taking the nucleon pairing into consideration
are shown in the Figs.~\ref{fig2}--\ref{fig4} (solid and dotted vertical lines,
respectively).

The coupling of the GT doorway states with many-quasiparticle configurations is
taken into consideration phenomenologically and is described in average over
the energy in terms of the smearing parameter $I(\omega)$ simulating the mean
spreading width of the doorway states. Following~[\cite{Moukh99}]
we choose $I(E_x)$ to be an universal function revealing saturation at rather
high excitation energies:
\beq
I(E_x)=\alpha\frac{E_x^2}{1+E_x^2/B^2},
\label{gams}
\eeq
where $\alpha$ and $B$ are adjustable parameters, $E_x$ is the excitation
energy in the daughter nucleus. Equation (\ref{gams}) is close to the
parametrization of the intensity of the optical-potential imaginary part
obtained within the modern version of an optical model. The choice of the
parameters $\alpha=0.09$ MeV$^{-1}$ and $B=7$ MeV has allowed us to describe
satisfactorily the total widths of a number of the isovector giant resonances
(including GTR) in the $^{208}$Pb parent nucleus~[\cite{Rod01}]. In this work
the energy-averaged GT strength function $\bar S_1^{(-)}$ is calculated as
$\bar S_1^{(-)}(\omega)= S_1^{(-)}(\omega+{\rm i}I(E_x)/2)$ using the same
parametrization of $I$, where $S_1^{(-)}$ is the GT strength function
calculated according to the CRPA or pn-QCRPA equations. The calculation results
are shown in the Figs.~\ref{fig2}--\ref{fig4} (thin line). The calculated
energy dependence in the vicinity of the GTR is approximated by the
Breit-Wigner formulae to get both the GTR energy $E_{GTR}$ and width
$\bar\Gamma_{GTR}$. The values of $E_{GTR}$ obtained thereby and
$\bar\Gamma_{GTR}$ along with the values calculated without taking the nucleon
pairing into consideration (in brackets) are listed in the Table~\ref{tab3} in
comparison with the corresponding experimental data.

We analyze the effect of the isospin splitting according
to Eqs.~(\ref{1.21}), (\ref{1.23})
without taking the nucleon pairing into consideration because this effect turns
out to be weak even for such a rather light parent nucleus as $^{90}$Zr.
In this case there is the only $T_0$ component of the GT strength function in
$^{90}$Nb (due to the $g_{9/2}\to g_{7/2}$ M1 transition in the proton
subsystem). Its relative strength and energy are $x_>=4.7$ \%
and $E_>=12.8$ MeV, respectively. It leads to a decrease in the relative
strength and energy of the GTR equal to $x_>$ and $x_>(E_>-E_<)/x_<=0.3$ MeV,
respectively, as compared with the values found within the CRPA.

\section{Discussion of the results}
\subsection{Summary concerned to the approach}
We have extended the CRPA approach, used previously
in~[\cite{Moukh99}--\cite{Rod01}] for describing the GTR and the IAR in
closed-shell nuclei, and have formulated a version of the pn-QCRPA approach
for describing the GT and F strength distributions in open-shell nuclei.
The common ingredients of both approaches are the following:
the phenomenological isoscalar part of the nuclear mean field;
the isovector part of the Landau-Migdal particle-hole interaction;
the isospin selfconsistency condition being used to calculate the symmetry
potential; mean Coulomb field calculated in the Hartree approximation;
the use of the full basis of the single-particle states in the particle-hole
channel and, as a result, formulating the continuum versions of the RPA;
a phenomenological description of the coupling of the particle-hole doorway
states with many-quasiparticle configurations in terms of the mean
doorway-state spreading width (the smearing parameter).

The specific feature of the developed pn-QCRPA approach is the use of an
isospin-invariant particle-particle interaction to describe both the nucleon
pairing phenomenon in neutral channels and particle-particle interaction in
charge-exchange channels. The method to check the isospin selfconsistency,
which could be violated by the use of a truncated basis of single-particle
states in particle-particle channel within the approach, is also used.

\subsection{F\,(C) strength distribution}
The present results (Table~\ref{tab2}) show at the microscopical level that the
isospin is a good quantum number for medium-heavy mass nuclei. In particular,
the IAR exhausts almost all F strength. The rest is mainly exhausted by the
IVMR. The heavier a nucleus is, the more the relative F strength of the IVMR
becomes. Similar results have also been obtained for the energy-weighted
Fermi-strength distribution.

The appropriate choice of the particle-particle interaction along with the use
of the same truncated basis of single-particle states in the neutral and
charge-exchange channels ensure the isospin selfconsistency of the approach
(the calculated values of $z$ are found to be less than $0.1\%$). For the same
reason, the effect of the nucleon pairing on the F strength distribution is
small.

The systematic lowing of the IAR energy compared with the experimental value is
a shortcoming of the approach, possibly caused by the absence of the full
selfconsistency.

\subsection{GT strength distribution}
\begin{figure}[ht]
\begin{center}
\epsfig{file=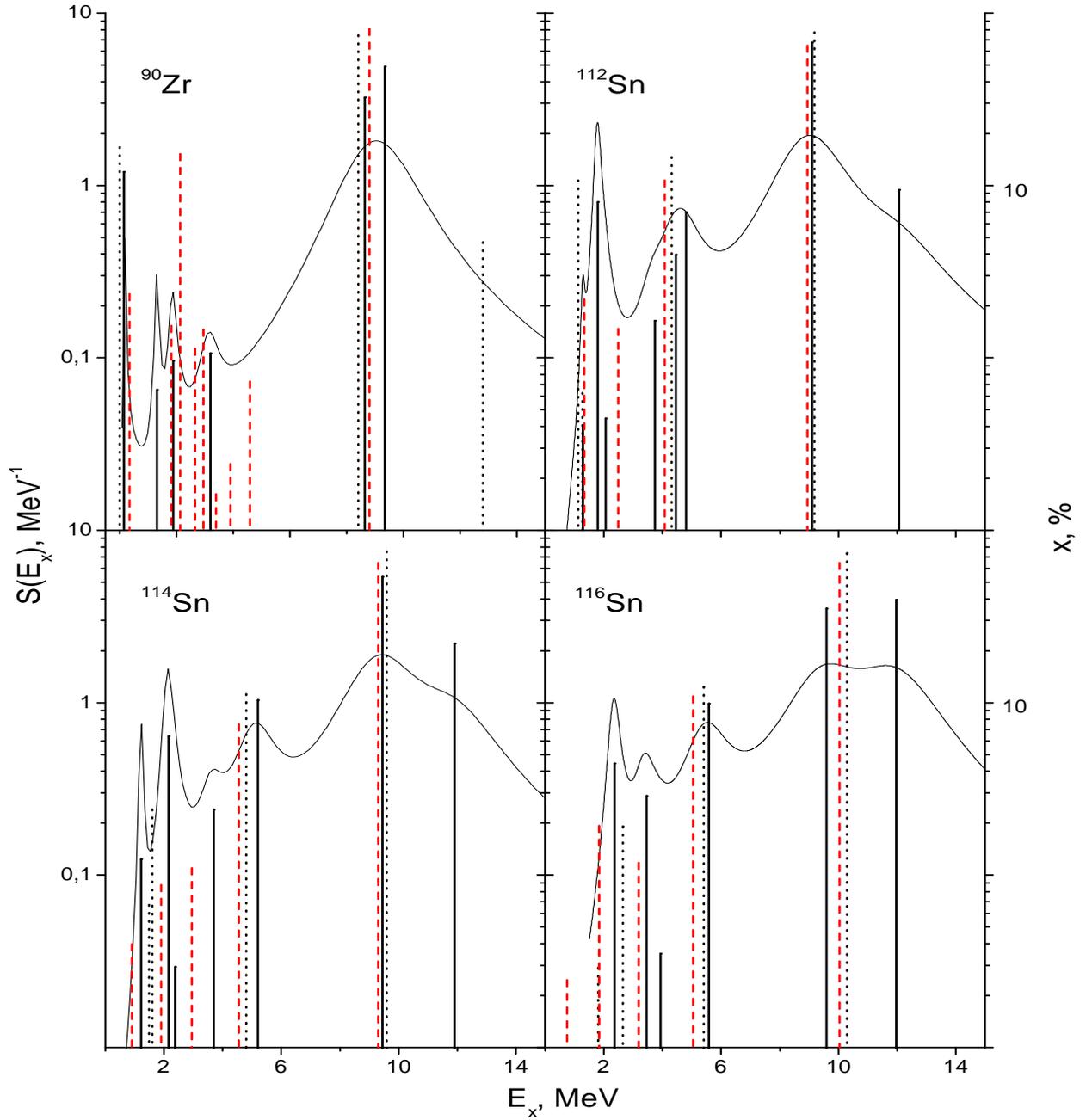,height=20cm,width=17cm}
\caption{The relative GT strengths $x$ in $^{90}$Nb, $^{112-116}$Sb calculated
within the pn-QCRPA (solid vertical line) and CRPA (dotted vertical lines)
in comparison with the respective experimental
data~[\protect\cite{Jan93,Kras01}] (dashed vertical lines).
The smeared pn-QCRPA GT-strength distribution $\bar S$ is also shown
(thin solid lines)}
\label{fig2}
\end{center}
\end{figure}

\begin{figure}[ht]
\begin{center}
\epsfig{file=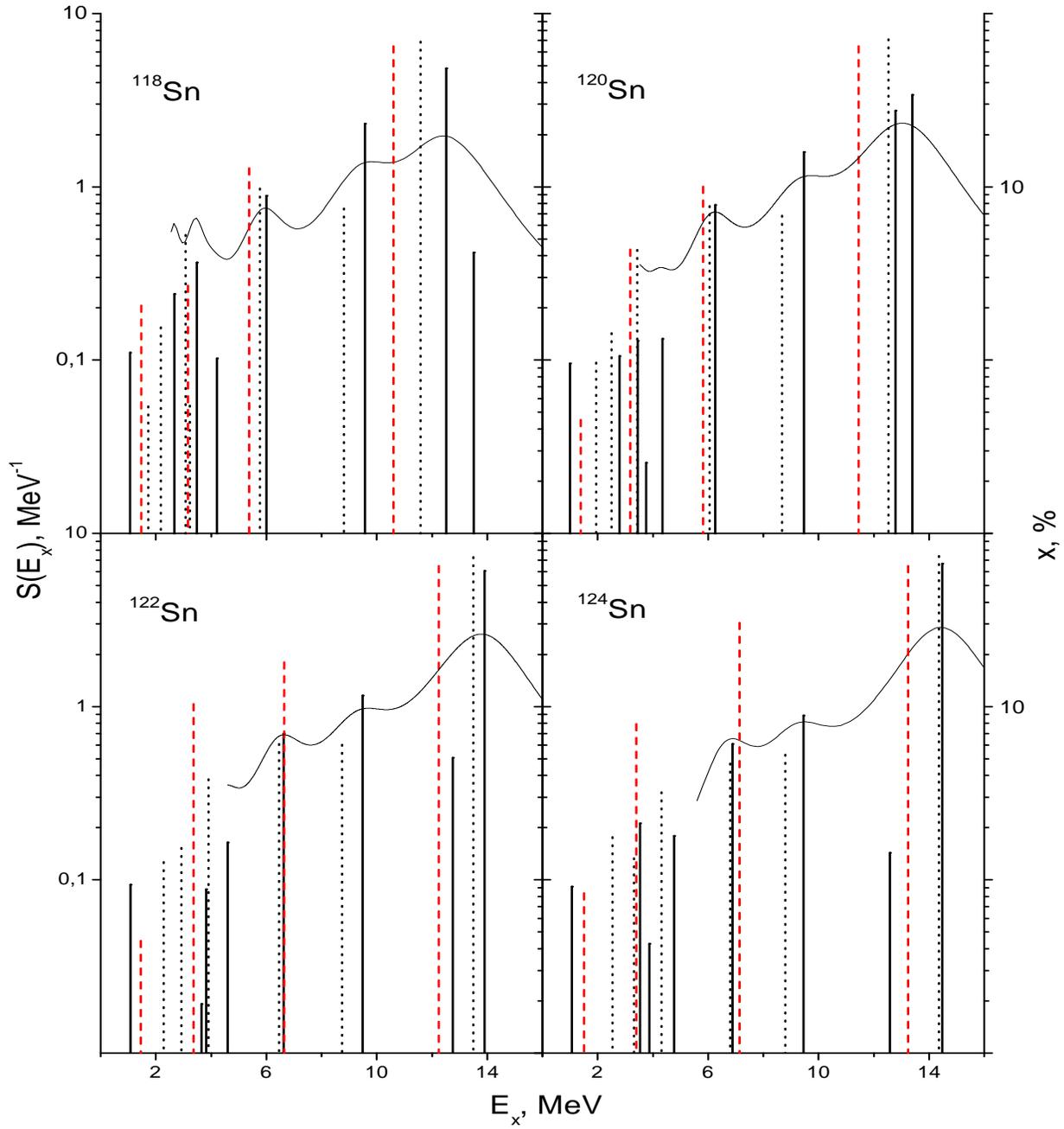,height=20cm,width=17cm}
\caption{The same as in the Fig.~\ref{fig2} but for the $^{118-124}$Sb (the
experimental data are taken from~[\protect\cite{Jan93}])}
\label{fig3}
\end{center}
\end{figure}

\begin{figure}[ht]
\begin{center}
\epsfig{file=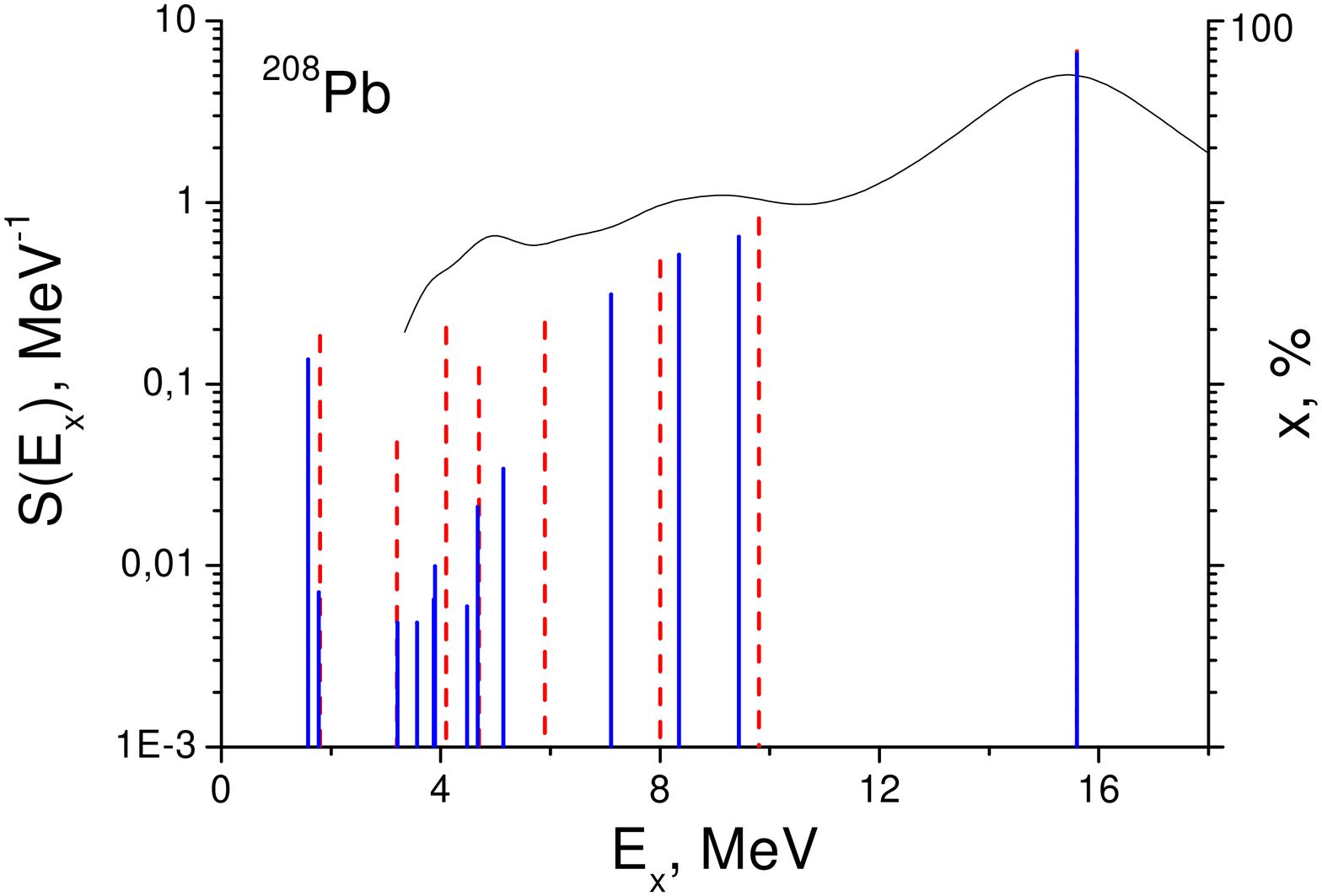,height=15cm,width=16cm}
\caption{The relative GT strengths $x$ in $^{208}$Bi calculated within the CRPA
(solid vertical lines) along with the respective smeared distribution
$\bar S$ (thin solid line). The experimental data are taken
from~[\protect\cite{Kras01}] (dashed vertical lines)}
\label{fig4}
\end{center}
\end{figure}

For all nuclei in question, the calculated GT strength function splits into
three main regions (see Table~\ref{tab3} and Figs.~\ref{fig2}--\ref{fig4}).
The main part of the GT strength (70--80\%)
is exhausted by the GTR. In the case of $^{208}$Bi, the calculated GTR relative
strength is found to be in a reasonable agreement with the respective
experimental value $x^{exp}_{GTR}=(60\pm 15)\%$~[\cite{akim95}].
The spin-quadrupole part of the particle-hole interaction is taken into account
only for $^{208}$Bi. As a result, the calculated GTR relative strength is
decreased from 68\% to 66\%
and the low-energy GT strength is noticeably redistributed. In the case of
$^{90}$Nb, the difference between experimental and calculated results is
somewhat larger: $x^{exp}_{GTR}=(39\pm 4)\%$~[\cite{Jan01}],
$x^{exp}_{GTR}=(61\pm 8)\%$ ($x^{exp}=(66^{+20}_{-10})\%$ up to 20 MeV
in excitation energy)~[\cite{Moos98}]. This could be partly due to the
well-known quenching effect which has been discussed during the last two
decades. This effect is not finally established experimentally well enough and
its discussion is out of the scope of this work.

The calculated high-energy part of the \GTSD is mainly exhausted by the
IVSMR and its satellites. This part is relatively large for $^{208}$Bi
(about 19\%)
and decreases up to about 9\%
for $^{90}$Nb (4.6\%
due to the IVSMR and 4.5\%
due to $T_>$ component of the GTR). The low-energy part contains the
weakly-collectivized GT states, which can be related to those found
in~[\cite{ga74,Guba89,VanG98}]. The nucleon pairing leads to noticeable
enriching of the calculated low-energy part of the \GTSD in open-shell
nuclei. This fact is in a qualitative agreement with respective experimental
data (shown in Figs.~\ref{fig2}--\ref{fig4}). The calculated values
$x_{low}/x_{GTR}$ 19.2\%
and 25.8\%
are found to agree reasonably with the corresponding experimental values 28.2\%
and 30.4\%
[\cite{Kras01}] for $^{90}$Nb and $^{208}$Bi, respectively. Bearing in mind
also the above-mentioned calculated and experimental values of $x_{GTR}$,
we can expect that the quenching effect is not so noticeable as it is usually
assumed.

The configurational splitting of the main GT state is found for some
single-open-shell nuclei (Figs.~\ref{fig2}, \ref{fig3}). In the case of the
$^{90}$Zr parent nucleus, the splitting is caused by the direct spin-flip
transition $1f_{7/2}^n\to 1f_{5/2}^p$, which becomes possible due to the proton
pairing and whose energy is close to the GTR energy calculated without taking
this transition into account. The value of the splitting (about 0.5 MeV) is
found to be rather small. For this reason, the total GTR width is only slightly
increased. In the case of the Sn isotopes, the strongest effect caused by the
$1h_{11/2}^n\to 1h_{9/2}^p$ transition is found for the $^{120}$Sn parent
nucleus. The splitting energy is found to be rather large and leads to a
noticeable increase of the total GTR width. The calculated isospin splitting
energy for $^{90}$Nb (4.0 MeV) is found to be in good agreement with the
respective experimental one (4.4 MeV~[\cite{Jan01}]).

In conclusion, we incorporate the BCS model in the CRPA method to formulate a
version of the partially self-consistent pn-QCRPA approach for describing the
multipole particle-hole strength distributions in open-shell nuclei.
The approach is applied to describe the Fermi and Gamow-Teller strength
functions in a wide excitation-energy interval for a number of single- and
double-closed-shell nuclei. A reasonable description of available experimental
data is obtained.

Authors are grateful to Prof.~M.~Fujiwara and Prof.~A.~Faessler for thorough
reading the manuscript and many valuable remarks. V.A.R. would like to thank
the Graduiertenkolleg ``Hadronen in Vakuum, in Kernen und Sternen'' (GRK683)
for supporting his stay in T\"ubingen and Prof.~A.~Faessler for hospitality.

\newpage
\begin{table}[ht]
\caption{Calculated values of the pairing gap $\Delta_\beta$ along with
the calculated and experimental  proton and neutron separation energies
$B_{p,n}^{calc,\ exp}$. The monopole pairing strength $G_0$ is also given}
\label{tab1}
\begin{center}
\begin{tabular}{|c|c|c|c|c|c|c|}
\hline
Nucleus & $G_0 \cdot A$, & $\Delta_\beta$, &$B_n^{calc}$, &
$B_p^{calc}$, &$B_n^{exp}$, &$B_p^{exp}$, \\
 & MeV &MeV  &MeV  &MeV  &MeV  &MeV  \\
\hline
$^{90}$Zr& 14.0& 1.15& 11.5&8.2& 11.98 & 8.36 \\
\hline
$^{112}$Sn& 11.5& 1.3& 10.9&7.5 & 10.79 & 7.56\\
\hline
$^{114}$Sn& 11.5& 1.4& 10.7&8.3& 10.30 & 8.48\\
\hline
$^{116}$Sn& 11.5& 1.3& 9.9& 9.1& 9.56 & 9.28\\
\hline
$^{118}$Sn & 11.5 & 1.4 & 9.7 & 9.9& 9.33 &10.00\\
\hline
$^{120}$Sn& 11.5& 1.4& 9.4& 10.7& 9.11 & 10.69\\
\hline
$^{122}$Sn& 11.5& 1.4& 9.1 &11.5 & 8.81 & 11.4\\
\hline
$^{124}$Sn & 11.5 & 1.2 & 8.7 & 12.3 &8.49 & 12.10\\
\hline
\end{tabular}
\end{center}
\end{table}

\begin{table}[ht]
\caption{Calculated characteristics of the Fermi strength distribution
along with respective experimental
data~[\protect\cite{Jan93,akim95,Kras01}] (see the text for details)}
\label{tab2}
\begin{center}
\begin{tabular}{|c|c|c|c|c|c|c|c|c|c|}
\hline
Parent &\multicolumn{3}{c|}{x, \%} & \multicolumn{3}{c|}{y, \% }&
\multicolumn{2}{c|}{$E_{IAR}$ , MeV} & $\bar E_{IVMR}$ , \\
\cline{2-9}
nucleus & IAR & high & (+)
& IAR & high & (+)
 & calc.& exp. & MeV\\
\hline
 $^{90}$Zr &
98.6 & 1.5 & 0.5 &
94.2 & 4.2 & 0.9 & 4.6 & 5.0 & 27.1\\
\hline
 $^{112}$Sn & 97.8 & 2.4 & 0.6 & 92.9 & 5.6 & 0.7 & 5.6 & 6.2& 25.0\\
\hline
 $^{114}$Sn & 98.0 & 2.4 & 0.5 & 93.5 & 5.5 & 0.7 & 6.8 & 7.3& 25.3\\
\hline
 $^{116}$Sn & 97.3 & 2.4 & 0.4 & 93.2 & 5.4 & 0.5 & 7.8 & 8.4& 25.5\\
\hline
 $^{118}$Sn & 97.7 & 2.4 & 0.3 & 93.9 & 5.3 & 0.5& 8.8 & 9.3& 26.0\\
\hline
 $^{120}$Sn & 97.6 & 2.4 & 0.3 & 93.9 & 5.2 & 0.4 & 9.7 & 10.2& 26.4 \\
\hline
 $^{122}$Sn & 97.5 & 2.4 & 0.2 & 94.0 & 5.2 & 0.3 & 10.7 & 11.2& 26.8\\
\hline
 $^{124}$Sn & 96.8 & 2.5 & 0.2 & 93.4 & 5.2 & 0.3 & 11.7& 12.2& 27.3 \\
\hline
 $^{208}$Pb & 94.4 & 3.1 & 0.2 & 89.8 & 6.2 & 0.2& 14.4 &15.1 & 32.5\\
\hline
\end{tabular}
\end{center}
\end{table}

\newpage
\oddsidemargin=-2.cm
\begin{table}[ht]
\caption{Calculated characteristics of the Gamow-Teller
strength distribution
along with respective experimental
data~[\protect\cite{Jan93,akim95,Kras01}] (see the text for details).}
\label{tab3}
\begin{center}
\hskip-2cm
\begin{tabular}{|c|c|c|c|c|c|c|c|c|c|c|c|c|}
\hline
Parent &\multicolumn{4}{|c}{x, \%} & \multicolumn{4}{|c}{y, \% }&
\multicolumn{2}{|c}{$E_{GTR}$, MeV} &
\multicolumn{2}{|c|}{$\bar\Gamma_{GTR}$, MeV}\\
\cline{2-13}
nucleus & low & GTR & high & ${(+)}$ & low & GTR & high & ${(+)}$ & calc.& exp.& calc.& exp. \\
\hline
$^{90}$Zr
& 14.8 &
76.5 & 9.1 & 4.1 & 6.7 & 79.3 & 8.2 & 2.4
&
8.8 (8.4)& 8.8$\pm$0.1 & 3.1 (2.7) & 4.8$\pm$0.2 \\
\hline
$^{112}$Sn
& 21.5 & 77.1 & 7.5 & 2.1 & 13.1 & 72.6 & 11.8 & 1.9
& 9.2 (9.2) & $8.94 \pm 0.25$ & 3.4 (2.8) & $4.8 \pm 0.3$ \\
\hline
$^{114}$Sn
& 20.6 & 75.6 & 6.9 & 3.8 & 12.4 & 72.8 & 11.3 & 1.6
& 9.9 (9.5) & $9.39 \pm 0.25$ & 4.4 (2.8) & $5.6 \pm 0.3$ \\
\hline
$^{116}$Sn
& 17.6 & 74.8 & 7.1 & 0.9 & 10.3 & 73.0 & 11.8 & 1.1
& 10.7 (10.3) & $10.04 \pm 0.25$ & 5.2 (3.0) & $5.5 \pm 0.3$ \\
\hline
$^{118}$Sn
& 17.1 & 75.6 & 7.0 & 2.0 & 9.3 & 74.1 & 11.6 & 1.3
& 11.9 (11.5) & $10.61 \pm 0.25$ & 4.9 (3.3)
& $5.7 \pm 0.3$ \\
\hline
$^{120}$Sn
& 28.6 & 61.4 & 7.8 & 1.2 & 19.5 & 63.4 & 12.8 & 1.3
& 13.0 (12.5) & $11.45 \pm 0.25$ & 4.5 (3.4) & $6.4 \pm 0.3$ \\
\hline
$^{122}$Sn
& 22.3 & 65.6 & 8.5 & 0.6 & 14.1 & 67.4 & 13.8 & 0.8
& 13.8 (13.5) & $12.25 \pm 0.25$ & 3.9 (3.5) & $5.6 \pm 0.3$ \\
\hline
$^{124}$Sn
& 21.6 & 66.9 & 9.0 & 0.6 & 12.5 & 68.5 & 14.3 & 0.8
& 14.4 (14.3) & $13.25 \pm 0.25$& 3.6 (3.5) & $5.2 \pm 0.3$ \\
\hline
$^{208}$Pb
& 17.0 & 65.8 & 18.7& 0.7 & 9.1 &  63.8 & 25.0 & 0.7
& 15.6& 15.6$\pm$0.2 & 3.7& 3.54$\pm$0.25 \\
\hline
\end{tabular}
\end{center}
\end{table}

\newpage
\oddsidemargin=-.5cm

\newpage

\newpage

\newpage

\end{document}